\documentclass{aa}
\begin{document}
   \thesaurus{06     
              (02.01.2;  
               02.08.1;  
               02.09.1;  
               02.16.1;  
               02.20.1;  
               09.03.1 
              09.11.1 
             09.03.1 
  10.03.1)} 

   \title{Angular momentum transport\\
in the central region of the Galaxy}


\author{Leonid M. Ozernoy$^{1,2}$, Alexei M. Fridman$^{3,4}$,
Peter L. Biermann$^5$}
          
   \offprints{L. Ozernoy}

   \institute{$^{1}$ Computational Sciences Institute and Department of Physics
\& Astronomy, George Mason U., Fairfax, VA 22030-4444, USA  \\
              email: ozernoy@science.gmu.edu\\
$^2$Laboratory for Astronomy and Solar Physics, Goddard Space Flight
Center, Greenbelt, MD 20771, USA\\
email: ozernoy@stars.gsfc.nasa.gov             \\
$^3$Institute of Astronomy, Russian Academy of Sciences,  48 Pyatnitskaya St.,
Moscow  109017, Russia\\
email: afridman@inasan.rssi.ru\\
$^4$Moscow State University, 18 University Prospect,
Moscow, 119899, Russia\\
$^5$Max Planck Institut f\"ur Radioastronomie, Auf dem H\"ugel 69,
P.O. Box 2024, D-53010, Germany\\
email: p165bie@mpifr-bonn.mpg.de
}
   
\date{Received ....; Accepted }

\authorrunning{L. Ozernoy et al.}
\titlerunning{Angular momentum transport}

\maketitle

\def\ul#1{$\underline{\smash{\vphantom{y}\hbox{#1}}}$}
\def\lax    {\ifmmode{_<\atop^{\sim}}\else{${_<\atop^{\sim}}$}\fi}
\def\gax    {\ifmmode{_>\atop^{\sim}}\else{${_>\atop^{\sim}}$}\fi}
\def\kms    {\ifmmode{{\rm ~km~s}^{-1}}\else{~km~s$^{-1}$}\fi}
\def\kmp    {~km~s$^{-1}$~pc$^{-1}$}
\def\lo     {~$L_{\odot}$}
\def\mo     {~${\rm M}_{\odot}$}
\def\moyr   {\hbox{~${\rm M}_{\odot}\,{\rm yr}^{-1}$}}
\def\etal   {{\sl et~al.~}}
\def\eg{{\it e.\thinspace g.~}}
\def\ie{{\it i.\thinspace e.~}}
\def\bk{\lower 6pt\hbox{${\buildrel k\over \sim}$}}
\def\bv{\lower 6pt\hbox{${\buildrel v\over \sim}$}}
\def\ref#1  {\noindent \hangindent=24.0pt \hangafter=1 {#1} \par}
\def\blankline  {\vskip12truept}


   \begin{abstract}

We discuss mechanisms for angular momentum transport in the clumpy
medium of the circumnuclear disk at the Galactic center. The viscosity due to
clump-clump collisions is found to be less than some critical viscosity;
this meets the conditions at which a collective mode of nonaxisymmetric
shear perturbations in the disk is able to grow
until going into the saturation regime where fully developed turbulence is
established. We find that the angular momentum transfer due to this
turbulent viscosity turns out to comparable to the transport due to magnetic
torques. Taken together, the turbulent and magnetic transfer of angular
momentum are able to provide the inflow of mass into the central parsec
with a rate of about $10^{-2}$\moyr, consistent with the available data.

   \end{abstract}

%

\section{Introduction}

Transport of angular momentum in the disks of spiral galaxies is one of
the central issues in galactic dynamics. It is also of prime
importance for fueling the central engines in active galactic nuclei (AGN).
With regard to our own Galaxy this topic has been recently explored in a
series of papers (von Linden et al. 1993a,b; Biermann et al. 1993) with
the conclusion that the energy input from the SN explosions can
feed the turbulence of the interstellar matter so as to provide  the
effective viscosity high as required to feed the star formation.

An immediate implication is that the viscous transport of angular
momentum outwards is accompanied by inflow of gas inward with the rate that
might be as high as $10^{-2}$\mo/yr on the scale of hundreds pc. Earlier, a
similar mass inflow rate was inferred to exist on the scale of the
circumnuclear disk, i.e. at 1.5 pc $\lax R\lax 10$ pc (Genzel \& Townes 1987,
Jackson et al. 1993).

We also note that the star formation in the inner region of the Galaxy
requires such a mass supply.

However, appropriate physical mechanism(s) which would
be able to provide  the necessary rate of momentum transfer within the
central 10 pc or so is (are) obscure so far. The present paper aims at
consideration of such mechanisms. In Sec. 2 and Appendix, we discuss
viscosity in a clumpy disk due to clump interactions, both with and without
self-gravity included. Sec. 3 deals with short-wave instability of the clumpy,
viscous disk; the necessary condition for developing small-scale
turbulence is established here. In Sec. 4, we apply this mechanism to the
circum-nuclear ring at the Galactic center. The results of the paper are
discussed in Sec. 5.


\section{~Viscosity due to cloud-cloud interactions}

In a recent version of the unified model for fueling AGN, Begelman et
al. (1989) proposed that the inflow of matter (driven by global
axisymmetric gravitational instabilities on large scales) proceeds on small
scales in the form of a `disk' composed of randomly moving clouds, which are
embedded in a low-density medium with a small filling factor. There is an
essential uncertainty in our
knowledge about the overall configuration, dynamics, and confinement
mechanisms for thermal gas clouds observed in AGN, but as for our Galaxy,
such a disk is known to exist as the circum-nuclear disk, or ring that is
rather clumpy, indeed (Jackson et al. 1993). Begelman et al. (1989) assumed
that the viscosity in the disk is provided by collisions between clouds.
Below, this collisional mechanism of viscosity is compared with some others.

We consider the following simple model of a
cloudy disk: The clouds are orbiting in an external gravitational field
and have some random peculiar velocities. It is assumed that the
clouds have a small filling factor and are embedded in a low-density medium
that provides a confinement of the clouds. In the differentially-rotating
cloudy disk, the angular momentum is transported due to cloud-cloud
interactions (which include, and are not just restricted to, collisions).
Let us address the shear viscosity associated with these interactions.

The most elaborated models to calculate the shear viscosity have been
considered by Goldreich \& Tremaine (1978) and Stewart \& Kaula (1980),
hereinafter referred to as GT and SK, correspondingly. GT considered
{\sl contact} inelastic collisions of {\sl non-gravitating} spherical
particles obeying an anisotropic distribution function, whereas SK
considered {\sl gravitational} (elastic) encounters of particles obeying
a Maxwellian distribution. In the both cases, as shown in Appendix A,
the viscosity coefficient can be represented in the form
$$
\nu ={\sigma _v^2\over\Omega}{A_i\tau\over B_i^2\tau^2 +1} ~. \eqno (2.1)
$$
Here $\sigma _v$ is the one-dimensional velocity dispersion,
$\Omega$ is the orbital angular
velocity in the disk, $\tau = \Omega t_i$ is the `optical depth'
to cloud-cloud interactions
(see below), $t_i$ is the free path time (index $i=c,~ G$; $c$ stands
for $c$ollisional, or $c$ontact, interactions, $G$ stands for $g$ravitational
ones), and $A_i,~B_i$ are the constant coefficients defined below.

Collisions between the clouds result in diminishing $\sigma _v$, whereas
gravitational encounters tend to increase it. For cloud-cloud collisions,
$$
t_c=\left(\pi a^2n\sigma_v\right)^{-1}, \eqno (2.2)
$$
where $a$ and $n$ are the typical size of a cloud and the spatial number
density of the clouds, respectively. For gravitational encounters
between the clouds
$$
t_G={3\sigma_v^3\over 4\sqrt\pi G^2m^2n} \eqno (2.3)
$$
(Braginskii 1965), where $m$ is the typical mass of a cloud. Evidently, $a_G=
2\left(3\sqrt\pi\right)^{-1/2}Gm/\sigma_v^2$ can be considered as an
effective size of the domain for gravitational influence of the cloud.

Coefficients $A_i$ and $B_i$ in Eq. (2.1) take the following values:
\begin{eqnarray}
A_c  &=&  0.46; ~~ \qquad \qquad  \qquad  \qquad  \quad   B_c  =  0.97  
\nonumber  \\
A_G  &=&  \left\{ \begin{array}{ll}
1.25 & {\rm if}~ v={\rm const}, \\ 
0.83 & {\rm if}~ \Omega \propto r^{-3/2}
\end{array}
\right.
;    \quad  B_G = 1.95.  ~~~~~~~~~~(2.4)  \nonumber  
\end{eqnarray} 

The optical depth $\tau$ of the disk is a convenient parameter describing
how effective are the interactions between the clouds.
By order of magnitude, it is nothing but the mean number of interactions
suffered by a cloud in passing through the disk. More accurately,
$$\tau \simeq \pi a^2nh~, \eqno (2.5)$$
assuming $a$ to be the largest of geometrical and gravitational influence
sizes. Here $h$ is the thickness of the disk given by
$$h\simeq {\sigma_v\over \Omega}~. \eqno (2.6)$$
Since $h\simeq\Sigma/mn$, where $\Sigma$ is the surface density of the disk,
Eq. (2.2) can be rewritten as
$$\tau\simeq {\pi a^2\over m/\Sigma}~, \eqno (2.7)$$
which implies one more interpretation for $\tau$: it is the covering factor,
$C$, or the fraction of disk area covered by clouds when they are placed as a
monolayer.

The filling factor of the system of clouds, \ie the
fractional volume filled by the clouds is
$$F={4\over 3}\pi a^3n~. \eqno (2.8)$$
\noindent
Eqs. (2.5), (2.6), and (2.8)
yield one more expression for $\tau$ containing $F$:
$$\tau\simeq {\pi a^2n\sigma_v\over \Omega}={3\over 4}F{h\over a}~.
\eqno (2.9)$$
Evidently, $F\ll 1$ for any cloudy disk with $a\ll h$ unless $\tau\gax h/a
\gg 1$.

Here we emphasize that cloud-cloud collisisons cannot be discussed without
noting that the magnetic fields permeate clouds, and are likely to make such
collisions more efficient by increasing the effective cross section:  When two
clouds collide, it is unlikely that they just slide along a given straight flux
tube.  First of all, if this imagined flux tube were not exactly on a circle,
then by angular momentum conservation the clouds would not go in a straight
line, and second, by virtue of the general distribution of velocities it is
very unlikely that two clouds would just match in proper velocities to be able
to slide along a flux tube, and, third, the energy density in flux tubes is
unlikely to sufficiently overpower the kinetic energy of clouds to do this. 
It follows that it is indeed likely that the flux tube will be twisted, thus 
strengthened in magnetic field, and therefore the clouds may interact
even at some distance.  This means that cloud-cloud collisions may involve a
larger effective cross section than just the geometry would imply.

Furthermore, with magnetic loops and reconnection in the region above the
disk, the effective scale height may well be larger than the scale height of
the visible cloud distribution.

It follows that the estimate above may be an underestimate just as well as an
overestimate; the observational fit and interpretation given to the data by
von Linden et al. suggests that the viscosity derived above for cloud-cloud
collisions is an underestimate.

It is instructive to compare the viscosity coefficient in a cloudy disk
[Eq. (2.1)] with that in a typical thin, but continuous disk:
$\nu\simeq v_sl$, where $v_s$ and $l$ are the sound velocity and the mean free
path length, correspondingly. Qualitatively, in a continuous disk
 $l$ is anticipated to be much smaller than $l$ in a cloudy disk. If $v_s\sim
 \sigma _v$, the viscosity in a cloudy disk exceeds that
in a continuous disk, bearing in mind some reasonable assumptions about the
disk parameters. Before specifying them,
we would like to discuss one more mechanism for viscosity in a cloudy disk,
this time of a collective origin, proposed recently by Fridman \&
Ozernoy (1992).


\section{~Small-scale turbulent viscosity}

 We consider, for simplicity, both the typical size of the clouds, $a$,
and the mean free path of the clouds, $l$, to obey an inequality:
$$a,~l\ll h~. \eqno (3.1)$$
The hydrodynamics of a cloudy accretion disk with
respect to 2-D nonaxisymmetric shear perturbations is similar to that
for a disk with the continuous distribution of the matter
explored  by Fridman (1989) who found the solution of this problem
in a small-amplitude, short-wave limit
$$k_rk_{\varphi}h^2 \gg 1, \eqno (3.2) $$
where $\vec k$ is the wave vector.
In a local rotating coordinate system,  the solution for the radial
component of the perturbed velocity, $v_{1r}$, reads:
$${v_{1r}(\tilde t)\over v_{1r}(0)}={1+\beta ^2\over 1+(\beta + \tilde t)^2}
\exp(-\nu/\nu _{\rm cr})~, \eqno (3.3)$$
where
$$\beta \equiv {k_r(0)
\over k_{\varphi}},~~\tilde t\equiv At,~~A\equiv -r{{\rm d}\Omega \over
{\rm d}r}~,     \eqno (3.4)$$

$$\nu _{\rm cr} \equiv A[k_{\varphi}^2\tilde t (1+\beta ^2 +\beta \tilde t +
{1\over 3}\tilde t ^2)]^{-1}~. \eqno (3.5) $$

\noindent
Here $\Omega (r)$ is the angular velocity of the disk, $t$ is the
time elapsed since the perturbations were ``turned on", and $\nu _{\rm cr}$
is a (time-dependent) critical viscosity. When $\nu\ll\nu _{\rm cr}$,
viscosity does not play any role in the disk dynamics.

The shortwave perturbations under consideration
 behave as incompressible modes. The solution
(3.3) describes how the vorticity decays with time in a viscous fluid. Some
limiting cases of interest can be revealed from Eqs. (3.3)--(3.5).
As $t\rightarrow \infty $, there appears the asymptotic solution
$\sim {\rm exp}(-\alpha t^3)$, which has been known earlier
(\eg Timofeev 1976, Zaslavskii \etal 1982).
Another limiting case when viscosity is absent ($\nu =0$), deserves a more
detailed consideration.


\subsection{ Non-viscous case ($\nu =0$)}

In the absence of viscosity, the solution (3.3) goes into
$$ v_{1r}(\tilde t)={v_{1r}(0)(1+\beta ^2)\over 1+(\beta +\tilde t )^2}~,
\eqno (3.6)$$
which was obtained by Lominadze \etal (1988). In this case, solution (3.6)
implies that
$$v_{1r}(\tilde t)k_{\perp}^2(\tilde t)=v_{1r}(0)k_{\perp}^2(0)={\rm
const}~, \eqno (3.7)$$
where
$$k_{\perp}^2(\tilde t)\equiv k_r^2(\tilde t) +  k_\varphi^2,~
k_r(\tilde t)=k_r(0) +  k_\varphi \tilde t~. $$
Eq. (3.7) is a 2-D analog of the Thomson theorem on the conservation of
vorticity in an incompressible fluid flow (Fridman 1989).

The solution (3.6) is shown in Fig.~1 by the solid lines. Evidently, the
velocity perturbation
is growing (implying instability) if the denominator in (3.6) is decreasing.
At the moment when the denominator has a minimum, i.e.
$\beta + \tilde t_* =0$, where $\tilde t_*$ is given by
$$\tilde t_* \equiv \left|{k_r(0)\over k_{\varphi}}\right|, \eqno (3.8)$$
$v_{1r}(\tilde t)$ reaches its maximum equal to
$$v_{1r~{\rm max}}(\tilde t_*)\simeq \left ({k_r(0)\over k_{\varphi}}
\right)^2v_{1r}(0) \eqno (3.9)$$
and then goes to zero at $\tilde t \rightarrow \infty$.
The growth of $v_{1r}$ at $0\leq \tilde t\leq -\beta$ is a result of
a decrease of $k_{\perp}(\tilde t)$ while $v_{1r}(\tilde t)k_{\perp}^2
(\tilde t)$ keeps constant due to the Thomson theorem (3.7).

The minimum of the denominator in Eq. (3.6) at $(\beta + \tilde t_*)=0$
implies that $\left(k_r(0)/k_\varphi+r\left|d\Omega/dr\right|t\right)=0$ as
$d\Omega/dr<0$. Since $r>0,~ t>0$, the necessary condition for the
perturbations to grow with time is given by
$${k_r(0)\over k_{\varphi}}<0~. \eqno (3.10)$$
Therefore, in a differentially rotating disk, the growing
short-scale spiral perturbations can only be leading.

To clarify possible misunderstandings, we note several points in the following:

In Appendix B, we show that in a differentially rotating disk there are
no incompressible short-wave perturbations oscillating with the
epicyclic frequency at all; in a solid-body rotating disk such perturbations
are stationary, in agreement with our eqs.(3.6)-(3.10).

In Appendix C, we
demonstrate that Coriolis forces are
irrelevant to the solution for velocity perturbations in a plane shear
layer, which we found to be similar to short-wave, low-frequency
perturbations in a rotating gaseous disk. We emphasize that Coriolis forces
have nothing to do with the
physical meaning of our solution. The latter is associated with
vorticity conservation (in the limit of small viscosity).

We consider perturbations with wavelengths {\it smaller} than the
disk's width. Consequently our case is close to a cylinder fluid layer.
Meanwhile the short-scale perturbations under the condition (3.2) are
incompressible (Fridman 1989). Therefore analytical results obtained in
this paper as well as in our previous papers (Fridman 1989,
Fridman and Ozernoy 1992) lend to support 
interpretation of well-known laboratory
experiments for a liquid flow between two rotating cylinders.
The latter demonstrate a nonlinear instability and developed
turbulence (e.g., Lukashchuk \& Predtechenskii 1984). 
Therefore, we have all the reasons to believe
that, for short-wave perturbations, a nonlinear
instability is likely to occur. Likewise, in Appendix 3, we show that
the dynamics of short-wave incompressible perturbations in a rotating disk
is completely equivalent to that in a plane shear layer.
This proves the equivalence of the
nonlinear instability in a plane shear layer
and in the situation investigated here.

It is instructive to compare the above instability with that
considered by Goldreich \& Lynden-Bell (1965) (referred to as GLB
hereinafter) for a self-gravitating, differentially rotating disk.
There are two basic differences between our and their situations.
First, different branches of growing perturbations are considered: we deal
with the vortex branch while GLB do with the sound-gravitational branch.
Second, the conditions for the growth of perturbations are different, too.
Under condition (3.10), the leading vortex perturbations grow,
due to the conservation of vorticity, in an otherwise stable disk.
GLB deal with the disk stable for axially-symmetrical perturbations,
and the growth of their non-axially-symmetrical, long-wave perturbations
proceeds due to a leading role of
self-gravity. Toomre (1982) christened this growth as ``swing
amplification".  It is worth noting that, in our case, a similar swing
amplification occurs for the (vortex)  short-wave perturbations and
it does not require self-gravity.


\subsection {General case (non-zero viscosity)}

The growth of perturbations with accounting for viscosity, which is described
by Eq. (3.3), is shown in Fig. 1 by the dotted and dashed lines.
The  ``instability" described by the exact solution (3.3) is of very
specific kind
as $v_{1r}$ tends to zero at $\tilde t \rightarrow \infty$. Nevertheless,
during a finite time $\tilde t_*$ given by Eq. (3.8) the perturbations are in
the growing regime and their amplitude increases by a factor of $k_{\perp}^2
(0)/k_{\perp}^2(-\beta)=(\beta^2+1)$, which can be $\gg1$ if $|\beta|=
|k_r(0)/k_\varphi|\gg 1$.

Let us suppose that by the moment $\tilde t =\tilde t_*\equiv \left|\beta
\right|$ when the
amplitude of any perturbation arrives at its maximum, the viscosity does not
play any essential role, \ie  $\nu\ll\nu _{\rm cr}(\tilde t_*)$. Though this
growth in the amplitude proceeds during the time interval
$\Delta \tilde t \simeq \tilde t_* $ only, it could onset the local
turbulence. In this case, turbulence can be established everywhere in the disk
as a superposition of the spiral perturbations originated in different
points of the disk and on different moments of time; all of them have
experienced a growth during $\Delta \tilde t \simeq \tilde t_*$.

The turbulence appeared as a result of the growth of the perturbations can
be characterized by some turbulent viscosity, which generally is much
larger than the molecular one. It is possible that a steady-state regime
will emerge in which
$$\nu \rightarrow\nu_{\rm cr}(\tilde t_*)\equiv \nu_{\rm turb}~. \eqno
(3.11)$$
In this regime, the decay due to viscosity is strong enough to provide a
steady-state level of the turbulence so that the amplitude of the
perturbations is kept more or less constant in time, \ie
$${v_{1r}(\tilde t)\over v_{1r}(0)}\simeq 1 ~~{\rm at}~~ \tilde t\gax
(\tilde t_*)= -\beta \simeq 1~. \eqno (3.12)$$

\noindent
Substituting Eqs. (3.11) and (3.12) into Eq. (3.3) with taking into account
Eq. (3.5) and assuming that, by order of magnitude, $A\simeq \Omega$,
we arrive at the following transcendental equation:

$${\rm exp}\left(-{4\over 3}~{\nu_{\rm turb}~k_{\varphi}^2\over \Omega}
\right)\simeq {1\over2}~. \eqno (3.13)$$

\noindent
Taking the logarithm of this equation one finds:

$$\nu_{\rm turb}\simeq 0.5~{\Omega\over k_{\varphi}^2}~. \eqno (3.14)$$

It is straightforward to see that the basic contribution into the turbulent
viscosity is given by the perturbations of the smallest $k_{\varphi}$'s;
therefore
$\nu_{\rm turb}\simeq 0.5\Omega/(k_{\varphi}^2)_{\rm min}$. From Eqs. (3.2)
and (3.12) one has $\left |k_{\varphi}\right |\simeq \left |k_r\right |\gg
h^{-1}$, whence $\left |k_{\varphi}\right |_{\rm min}\simeq h^{-1}$, \ie

$$\nu_{\rm turb}\simeq 0.5~\Omega h^2~. \eqno (3.15)$$
By substituting $h=\sigma_v/\Omega$ into Eq. (3.15) the latter can be
rewritten in the form:
$$\nu_{\rm turb}\simeq 0.5~{\sigma_v^2\over \Omega}~. \eqno (3.16)$$

It is instructive to compare the expression for the Bohm diffusion coefficient
$(D_B)_{\rm}$ with our Eqs. (3.14) and (3.15)
 for $\nu_{\rm turb}.$  A well-known estimation of the Bohm diffusion
coefficient for a strong turbulence plasma is given by
(see e.g., Kadomtsev, 1964, or a review by Horton, 1984):
$$
 (D_B)_{\rm max} \simeq {(\gamma_L)_{\rm max}\over (k^2_\bot)_{\rm min}},
$$
 where $\gamma_L$ is the maximum linear growth rate of the drift
 instability, and $(k_\bot)$ is the minimum wave number, also from a linear
 theory. They are much the same: in our case $(\gamma_L)_{\rm
 max} \sim
 0.5~ \Omega$ and instead of $k_\bot$ we substitute $(k_\varphi)_{\rm min}
 \simeq h^{-1}.$
 Kadomtsev obtained  $(D_B)_{\rm max}$ by using
 the relationship $k_\bot \rho \sim 1,$ where $\rho$ is the Larmor radius. At
 $k_\bot \sim \rho^{-1}$, $D_B$ has a maximum. But the Larmor radius in
 plasma corresponds to the epicyclic radius in graviphysics. The latter, in
 fact, is the thickness of disk, $k_\bot \simeq h^{-1}$, and this, actually,
 is being used here.

This viscosity whose value is given by Eq. (3.15) or (3.16) was called by
Fridman \& Ozernoy (1992) {\it anomalous} in the same
sense as one introduces anomalous resistivity and anomalous diffusion in plasma
physics: The origin of this viscosity is in fully developed turbulence
which is established in the saturation regime described above.

Eq. (3.15) or (3.16) could be derived from dimensional
arguments (of course, without the numerical coefficient)  as an estimation
of viscosity in a rotating disk with turbulent motions. However, we should
emphasize that without an analysis such as one given above it would be
impossible to reveal an
underlying physical mechanism for the origin of such turbulence.

It is important to test whether the necessary condition for developing
of small-scale turbulence
$$\nu < \nu_{\rm cr}   \eqno (3.17)$$
is met. To this end, let us find the ratio $\nu/\nu_{\rm turb}$ as a function
of $\tau=\Omega t_i$, \ie of the basic parameter that characterizes
the number of interactions per one revolution:
$${\nu\over \nu_{\rm cr}} \simeq{2A_i\tau\over B_i^2\tau^2 +1}~.
\eqno (3.18)$$
\noindent
Asymptotically,
\[
{\nu\over\nu_{\rm cr}}\simeq 
\left\{  \begin{array}{ll}
~~~2A_i\tau  \ll 1 & {\rm if}~ \tau\ll 1  ~, \\
{2A_i\over B_i^2}\tau^{-1}  \ll 1 &  {\rm if}
~\tau\gg 1  ~, ~~~~~~ (3.19)
\end{array}
\right.
\]
\noindent
\ie  in both limits, $\tau\ll 1$ and $\tau \gg 1$,
the viscosity coefficient is much less than the critical value given by Eqs.
(3.15) or (3.16). The function (3.18) reaches its maximum at $\tau
=B_i^{-1}$, and this maximum is given by
$$
\left ({\nu \over\nu_{\rm cr} } \right)_{\rm max} \simeq {A_i\over
B_i}~.  \eqno (3.20)
$$
Even in this, the least favorable case [when $\tau \simeq 0.5$ as one can
see from Eq. (2.4)]
$\nu$ is less than $\nu_{\rm cr}$ by a factor of 2 or so, which is enough
for small-scale turbulence to appear. Therefore the range
of physical conditions under which the viscosity due to fully developed
turbulence should dominate is indeed very broad.

Unlike the classical example of gravitational instability, our
mechanism for the growth of shortwave perturbations has a much higher
level of viscous stabilization, as it follows from the value of critical
viscosity calculated above in comparison with that for a
self-gravitating disk. Indeed, the growth of shortwave perturbations
leads to the increase of the amplitude by a factor of

$$
 {v_{1r}(t_*)\over v_{1r}(0)} \simeq ({k_r(0)\over k_\varphi})^2>>1.
$$

This strong inequality follows from the fact that
 $|k_r(0)/ k_\varphi| \sim t_*/T,$ where $t_*$ is the
characteristic time of the growth of perturbations and $T$ is the period
of the disk revolution. According to the perturbation theory implemented
to examine instability, the condition $t_*/T >> 1$ holds (otherwise
the zero-approximation of perturbation theory is not fulfilled:
the equilibrium condition is broken for the time less then that
of one revolution of the disk). As a result of the above inequality, a strong
growth of perturbation takes place, which leads to the development
of short-scale turbulence and the appearance of turbulent viscosity.
A very large factor of the growth given above explains why
the value of the critical turbulent viscosity, which stops the growth of
perturbations, turns out to be much larger than that 
for the instability of a self-gravitating disk (e.g., Fridman \& Polyachenko
 1984,  p. 41).


\section{~Viscosity in the Circumnuclear Ring}

Before making numerical estimates, we list  the basic parameters of the
clumpy gas in the circumnuclear ring (CNR), such as the inferred clump size
$a$, the volume filling factor $F$, and velocity dispersion of the clumps
$\sigma_v$, taken from Jackson et al. (1993) and G\"usten at al. (1987):

$$a\simeq 0.15~{\rm pc}; ~~~F\sim 0.1~-~0.3; ~~~ \sigma_v\simeq 20~{\rm
km~s}^{-1}~ . \eqno (4.1)$$

Adopting the average gas density in the clumps $n=10^5~{\rm cm^{-3}}$
one finds the average clump mass $m=5$\mo. This gives the ratio
$a_G/a\simeq 10^{-3}$, which implies that elastic (gravitational)
interactions between the clumps are negligible compared to inelastic ones,
\ie clump-clump collisions. In other words, gravitation plays no role in the
interactions between the CNR clumps.

The mean free path of the clumps given by
$$l={4\over 3}{a\over F}\simeq (0.7~-~2)~{\rm pc}~, \eqno (4.2) $$
\noindent
is rather large (even in a marginal conflict with the simplifying
assumption (3.1) that $l\ll h$). The clump-clump collision rate in the
CNR is given by:

$$\omega_c\simeq {\sigma_v\over \l}\simeq (0.3~-~1)\cdot 10^{-12}~
{\rm s}^{-1}~, \eqno (4.3)$$
\noindent
\ie $\omega_c\lax\Omega\simeq 2\cdot 10^{-12}~{\rm s}^{-1}$ implying
less than one collision per revolution. The anticipated optical
depth is $\tau\simeq 0.1~-~0.5$, which, according to Eq. (3.17), results
in $\nu < \nu_{\rm cr}$. Therefore, the conditions for fully developed
turbulence to appear, which are described at the end of Sec. 3, are met
to yield the viscosity coefficient

$$\nu_{\rm turb}\simeq 1\cdot 10^{24}~{\rm cm}^2{\rm s}^{-1}~.
\eqno (4.4)$$

It is instructive to compare this result with the upper limit to viscosity
derived by von Linden et al. (1993a,b). Their
results were obtained by fitting an accretion disk model
with arbitrary viscosity to the velocity fields of various molecular clouds,
and then inferring the required kinematic viscosity from the fit.  Successful
fits  were made in the radial range from 10 to 100 pc, with an implied
kinematic viscosity of $6 \, 10^{26} \, \rm cm^2 \, sec^{-1}$ at a distance of
100 pc.  Such a high viscosity is just within the limits imposed by the basic
assumption of an accretion disk:  The thin disk assumption implies, in the
context of isotropic turbulence, that the kinematic viscosity has to be
clearly less than the circular velocity times the scale height.  At $r\sim
100$ pc the circular velocity is $\sim 200$ km/sec, and the scale height is
difficult to determine; the $z$-distribution of clouds gives a lower limit to
the scale height, and that is $\sim 10$ pc.  This means that the kinematic
viscosity has to be less than $6 \, 10^{26} \, \rm cm^2 \, sec^{-1}$; the fit
by Linden et al. is obviously just at the limit.  It follows either, a) that
the real scale height is quite a bit larger, with a rather hard limit at
roughly 1/3 of the radius, implying a hard limit of the viscosity of $2 \,
10^{27} \, \rm cm^2 \, sec^{-1}$, which should not be reached, or b) that the
kinematic viscosity cannot be described with an isotropic turbulence.

The viscosity values implied by the fit to the observations of molecular clouds
differ for different radii.  A fit was made for clouds at 100 pc as well as 10
pc, with the viscosity decreasing for smaller radii. This radial variation may
become steeper at smaller radii.  The hard upper limit mentioned above would
imply that it decreases as approximately $r^{1.13}$, and so would imply
that the hard upper limit at 1.5 pc is $\approx 5 \, 10^{24} \, \rm cm^2 \,
sec^{-1}$.
We note that the result (4.4) is a factor of 5 below this upper limit.
Assuming the same scaling for the kinematic viscosity derived by von Linden
et al. would lead to $\approx 1.5 \, 10^{24} \, \rm cm^2 \, sec^{-1}$, which
is close to Eq. (4.4); assuming the number derived from a fit at 10 pc we
obtain an estimate which is near the limit.

\section{~Discussion}
\bigskip

A qualitative argument presented at the end of Sec. 2 shows that the
viscosity in a cloudy disk is, in general, much larger than that for a
continuous disk. This assertion could be easily confirmed by numerical
estimates when we address the circum-nuclear ring (CNR): in a continuous
disk with similar global parameters, one would have $\nu\simeq c_sl\lax c_sh
\simeq 10^{23}~{\rm cm}^2{\rm s}^{-1}$, where $c_s\simeq 1$ km/s is the
sound speed of the gas (G\"usten et al. 1987) and $h\simeq 0.5~{\rm pc}$ is
the thickness of the disk. What provides a much
larger viscosity than it would be possible in a continuous disk
is (although only partly) the clump-clump collisions.
As is shown in Sec. 3, the collective mode of instability would
dominate the dynamics of a clumpy disk if viscosity in the latter is less
than some critical viscosity [Eq. (3.11)] whose value is given by Eqs. (3.15)
or (3.16). This condition is (marginally) met in the CNR,
which results in the turbulent viscosity coefficient
$\nu_{\rm turb}\simeq 1~ 10^{24}~{\rm cm}^2{\rm s}^{-1}$.
We note in passing that G\"usten et al. (1987) suspected the existence
of turbulence in the clumpy CNR, although they did not evaluate its
viscosity coefficient.

It is worth mentioning that, in spite of the differences in the
specific physical parameters of the clumps in the CNR and active
galactic nuclei, or AGN (for the
latter, see Netzer 1990), the value of $\nu_{\rm turb}$ evaluated above
turns out to be in the range of the values estimated by Fridman \& Ozernoy
(1992) for cloudy disks in AGN. This might be relevant to the issue
whether the CNR could be considered as a prototype for circum-nuclear
tori around some types of AGN.
(It is interesting that a large value of the viscosity in the CNR implies
its rather large scale height, which is a required geometry for the AGN tori).
In any case, the angular momentum
transport in the CNR seems to be a template while considering similar
issues both for quiescent and active galactic nuclei.

The analysis performed above has not accounted for the magnetic field in
the CNR. Meanwhile several observational techniques revealed the field
strength in the clouds to be $\sim 1$ mG (for a recent review, see
Genzel et al. 1994). A magnetic field as strong as this cannot be ignored in
the transport of angular momentum. Conservation of angular momentum
transported by both viscous and magnetic stresses can be written for the
CNR in which $v_{\varphi}\approx$ const in the form (Ozernoy \& Genzel 1998):
$$\dot M= {2\pi r\Sigma\over v_{\varphi}}\left (\xi v_A^2 + \nu_{\rm
eff} \Omega \right
)\left[1-\left({R_i\over r}\right )^{1/2}\right ]^{-1}, \eqno (5.1)$$
where $\dot M$ is the mass inflow rate, $\Sigma$ is the surface density
of the CNR, $\xi \sim 1$ is the $(-B_r/B_{\varphi})$ averaged
over the $z$-coordinate, $v_A$ is the Alfv\'en velocity, $\nu_{\rm eff}$ is an
effective viscosity, and $R_i\simeq 1.5$ pc is the inner radius of the ring.
If we consider the CNR as a magnetized disk for which
$v_A\simeq h\Omega \simeq 30$ km/s  and $\nu_{\rm eff}$ is given by Eq. (4.4)
the two terms in parentheses in the r.h.s. of Eq. (5.1) turn out
to be comparable. This implies that while evaluating the angular momentum
transport in the CNR, both magnetic and turbulent
viscosity need to be accounted for. One can see that with the parameters
listed above and $\Sigma\simeq 2~ 10^{-2}$ g cm$^{-2}$ the total mass
inflow rate given by Eq. (5.1) at a fiducial distance of $r=2$ pc amounts
to $\dot M\simeq 10^{-2}$\moyr.

This result is consistent with a naive, by
order-of-magnitude, estimate of $\dot M\sim
M/t_{tr}$, where $M$ is the CNR mass and $t_{tr}=R_i^2/\nu_{\rm turb}$ is the
characteristic time for the angular momentum transport: Taking
$\nu_{\rm turb}\simeq 1~ 10^{24}~{\rm cm}^2{\rm s}^{-1}$ and $M\sim 3~
10^4$\mo, i.e. somewhere in between $10^4$ and $10^5$\mo, the current
estimates for the CNR mass (Genzel et al. 1995), one gets $t_{tr}\simeq
6~ 10^5$ yr and  $\dot M\sim 5~ 10^{-2}$\moyr.  This is consistent
as well with the inflow rate
toward the Galactic center inferred from the observational data (e.g. Blitz
et al. 1993, Genzel et al. 1994).

\blankline

\noindent{\bf Appendix A:} 

\noindent{\bf Shear Viscosity in a Clumpy Disk at Elastic/Inelastic Encounters}

The shear viscosity in a differentially rotating clumpy disk, with accounting
for gravitational interactions between the clumps of mass $m$, is given by
Eq. (60) in Stewart \& Kaula (1980):
\begin{eqnarray*}
-r\Omega^ \prime\nu &=&A_0\sigma_v^2, \\
A_0 &=&{9\sqrt 2~ \Omega\omega_G^{-1}\over (36/5)+55\left(\Omega\omega_G^{-1}
\right)^2}~, \qquad ~~~~~~(A1)
\end{eqnarray*} 
where $\Omega^\prime \equiv {\rm d}\Omega/{\rm d}r$, and $\omega_G$ is
frequency of clump-clump collisions given by
$$\omega_G={3.35~ G^2m^2n\over \sigma_v^3}~. \eqno (A2)$$
\noindent
Eq. (A2) corresponds to Eq. (57) of SK when one puts $\ln [1+(\sigma_v^3/
Gm\Omega)^2]^{1/2}=1$ because in a flat disk the long-range gravitational
interactions could be neglected compared to the short-range ones. The value
of $\omega_G$ is related to the characteristic time of gravitational
encounters, $t_G$, given by Eq. (2.3) simply by $\omega_G^{-1}=t_G/\sqrt 2$.
By substituting Eq. (A2) into Eq. (A1), the latter could be written in the
form:
$$\nu={\sigma_v^2\over \Omega}{A_G\tau\over \left(1.95~\tau\right)^2+1}~,
\eqno (A3)$$
where $A_G=0.83$ for a Keplerian disk, $A_G=1.25$ for a solid-body
rotating disk, and $\tau=\Omega t_G$.

The viscosity coefficient for clump-clump collisions, when the clump gravity
is negligible [Goldreich \& Tremaine 1978, Eq. (46)],
can be written in the form analogous to Eq. (A3):
$$\nu={\sigma_v^2\over \Omega}{0.46~\tau\over \left(0.97~\tau\right)^2+
1}~.  \eqno (A4)$$

Eq. (2.1) unifies the representation of viscosity both in the case when
interactions are inelastic, while gravitation is negligible [Eq. (A4)]
{\it and} in the case when gravitational encounters are dominating [Eq. (A3)].

\blankline

\noindent{\bf Appendix B:} 

\noindent{\bf Character of Perturbations in a 
Differentially Rotating vs. a Solid-body Rotating Disk}

Let us consider an incompressible, differentially rotating, liquid
cylinder. The linearized equations of motion take the form:
\begin{eqnarray*} 
{\rm div}\tilde{ \vec v} &= & 0, \\
{{\partial \tilde v_r}\over{\partial t}} ~+~\Omega {{\partial
   \tilde v_r}\over{\partial \varphi}}~-~2\Omega \tilde
   v_\varphi & = & -{{1}\over\rho} {{\partial \tilde P}\over{\partial r}}, \\ 
{{\partial \tilde v_\varphi}\over{\partial t}} ~+~\Omega
{{\partial \tilde v_\varphi}\over{\partial
\varphi}}~+~{{\kappa^2}\over{2\Omega}} \tilde v_r & = & -{{1}\over{\rho r}}
{{\partial \tilde P}\over{\partial \varphi}}. 
\end{eqnarray*} 

Substituting a perturbation in the form of the wave packet $f$ $\sim$
$\exp[-i\omega+m \varphi+\int k_r dr]$  results in the following dispersion
relation ($k_\varphi\equiv m/r$):
$$ 
\hat \omega ~\equiv~ \omega - m\Omega ~=~ -i (2\Omega -
\kappa^2/2\Omega) {{k_r k_\varphi}\over{k^2}}. 
$$
As always $\kappa^2$ $\le$ $4\Omega^2,$ and therefore the condition of
instability is $k_r k_\varphi$ $<$ 0, which corresponds to the growth
of the leading spirals. Under this condition the instability is monotonous
without any oscillations
with epicyclic frequency. If the rotation is solid-body, one
obtains  $\hat \omega$ = 0, i.e. the perturbations are stationary.

The above solution has a drawback since the $r$-component
of the group velocity for perturbations under consideration is zero. A more
detailed analysis should be performed to find out what is the role of
$r$-dependence for $\Omega$.
This difficulty can be overcome if one considers the system in a
corotating frame of reference. In this case, an exact time-dependent
solution can be derived (see Fridman, 1989) which
demonstrates a qualitatively the same behaviour as is obtained above.
Specifically, in any differentially rotating disk whose angular velocity
decreases with radius, the leading spirals grow monotonously.
As for a solid-body rotating disk, all perturbations are stationary.

Note that if one would consider absolutely compressible case, i.e.
$P=0$ in the above equations, the continuity equation takes the
form:
$$
{{\partial \tilde \rho}\over{\partial t}}~+~\rho {\rm div}\tilde{
\vec v}~+~v_r{{d\rho}\over{dr}}~=~0, 
$$
and we immediately obtain
$$ 
\hat \omega^2 ~=~ \kappa^2.
$$

The contrast to other work is illuminating.  Julian \&
Toomre (1966) considered perturbations in a very thin collisionless
self-gravitating stellar disk. In fact, for such perturbations
a collisionless {\it stellar} disk is absolutely compressible, in contrast
to incompressible character of perturbations with wavelengths
much smaller than the disk's width for a {\it gaseous} disk considered here.
Goldreich \& Lynden-Bell (1965) studied dynamics of perturbations in a gaseous
disk, but only at high frequencies. Really, perturbations
considered in their paper demonstrate (asymptoticaly) oscillations
with sound frequency $kc$, where $k$ is the wave number and $c$ is the sound
speed. As the scale of perturbations is smaller than the disk
width, i.e. $kh>1$, their frequency is higher than the epicyclic one. But
as the equations of motions contain three derivatives with respect to
time, they describe three modes of perturbations. Besides the two
sonic high frequency modes, one mode of low frequency perturbations exists
as well. What is examined in our paper is the role of {\it low frequency}
mode in dynamics of the circumnuclear disk. As was shown in Fridman (1989), 
this mode does not demonstrate any oscillations and is stationary in a 
solid-body rotating disk.

\blankline

\noindent{\bf Appendix C:}

\noindent{\bf Velocity Perturbations in a Plane Shear Layer}

We start from linearized equations for short-wave, low-frequency
perturbations in a barotropic gaseous disk in the corotating frame of
reference (for derivation, see Fridman 1989):
\begin{eqnarray*} 
k_r(t)\tilde v_r~+~k_\varphi \tilde v_\varphi & = & 0,\\
{{\partial \tilde v_r}\over{\partial t}} ~-~2\Omega_0 \tilde
v_\varphi & = & - i k_r(t)\tilde \chi - \nu k_\perp^2(t) \tilde v_r, \\
{{\partial \tilde v_\varphi}\over{\partial t}} ~+~(2\Omega_0-A) \tilde
v_r & = & - i k_\varphi\tilde \chi - \nu k_\perp^2(t) \tilde v_\varphi. 
\end{eqnarray*} 
Here $ \Omega_0$ = const is the angular velocity of the corotating frame of
reference; $A$ $\equiv$ $-r_0(d\Omega/dr)_0$  characterizes
the value of the shear; $k_r(t)$ $\equiv$ $k_r$ $+$ $tAk_\varphi$;
$k_\perp^2(t)$ $=$ $k_r^2(t)$ $+$ $k_\varphi^2$.

This system has an exact solution in the form (Fridman, 1989):
$$
\tilde v_r(t)~=~{{k_\perp^2(0)}\over{k_\perp^2(t)}} \tilde v_r(0)
\exp\left[-\nu\int\limits_0^t k_\perp^2(t)dt\right].  \eqno (C1)
$$
For small $t$ the exponential term is negligent and we come up with
vorticity conservation:
$$ 
{\rm curl}_z(\vec v)~=~i(k_rv_\varphi-k_\varphi v_r)~=~-{{ i}\over
{k_\varphi}} k_\perp^2 v_r~=~{\rm const}.  \eqno (C2)
$$

For a plane shear layer, a similar set of equations can be derived. The
simpliest way to do this would be to put in the above system
$\Omega$ $\rightarrow$ $0$, $r$ $\rightarrow$ $\infty$,
$\Omega r$ $\rightarrow$ $v(x)$. As a result, we obtain the following system:
\begin{eqnarray*} 
k_x(t)\tilde v_x~+~k_y \tilde v_y & = & 0,\\
{{\partial \tilde v_x}\over{\partial t}} & = & - i k_x(t)\tilde \chi -
\nu k_\perp^2(t) \tilde v_x, \\
{{\partial \tilde v_y}\over{\partial t}} ~+~B \tilde
v_x & = & - i k_y\tilde \chi - \nu k_\perp^2(t) \tilde v_y. 
\end{eqnarray*} 
Here $B$ $\equiv$ $(dv/dx)_0$ characterizes the
value of the shear; $k_x(t)$ $\equiv$ $k_x$ $-$ $tBk_y$;
$k_\perp^2(t)$ $=$ $k_x^2(t)$ $+$ $k_y^2$. One can easily see
that this system is almost similar to the previous one and it has a
similar solution:
$$
\tilde v_x(t)~=~{{k_\perp^2(0)}\over{k_\perp^2(t)}} \tilde v_x(0)
\exp\left[-\nu\int\limits_0^t k_\perp^2(t)dt\right].  \eqno (C3)
$$
Obviously, in a plane layer there are no any Coriolis
forces at all. Nevertheless, the solution (C3) has the same form
as (C1). 
What is relevant
to the real physical meaning of our solution is conservation of vorticity
(eq. C2), which has been emphasized here.

\def\apj    {{ApJ{\rm,}\ }}
\def\apjl   {{ApJ~(Letters){\rm,}\ }}
\def\apjs   {{ApJ~Suppl.{\rm,}\ }}
\def\aa     {{A\&A{\rm,}\ }}
\def\anrev  {{Ann.~Rev.~Astr.~Ap.{\rm,}\ }}
\def\baas   {{Bull.~AAS{\rm,}\ }}
\def\mnras  {{M.N.R.A.S.{\rm,}\ }}
\def\ref#1  {\noindent \hangindent=24.0pt \hangafter=1 {#1} \par}
\def\v#1  {{{#1}{\rm,}\ }}


{\bf Fig. 1.} ``Unstable" solutions, $y\equiv v_{1r}(\tilde t)/ v_{1r}(0)$,
 represented by Eq. (3.3) for several
values of $\beta\equiv k_r(0)/k_\varphi$. The solid lines show a non-viscous
case  (Eq. 3.6). A general, non-zero viscosity solution (3.3) is
shown by the dotted lines for $\nu=0.3~\nu_{\rm crit}$ and by the dashed
lines for $\nu=\nu_{\rm crit}$, where $\nu_{\rm crit}$ is defined by Eq.
(3.5).

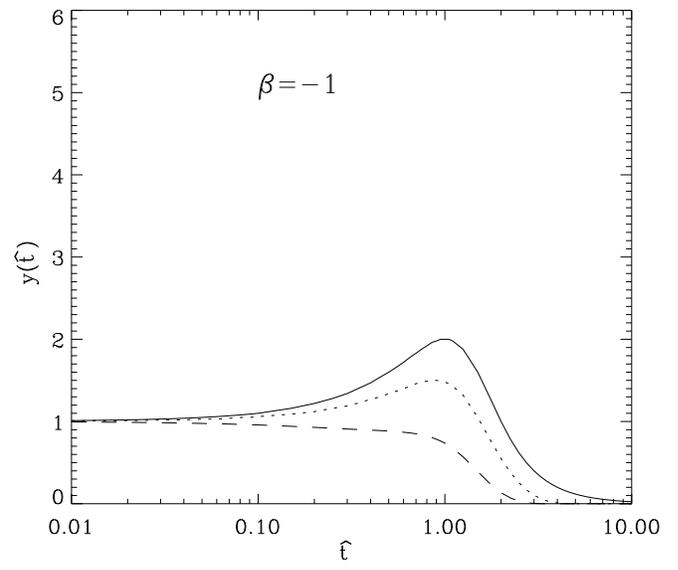
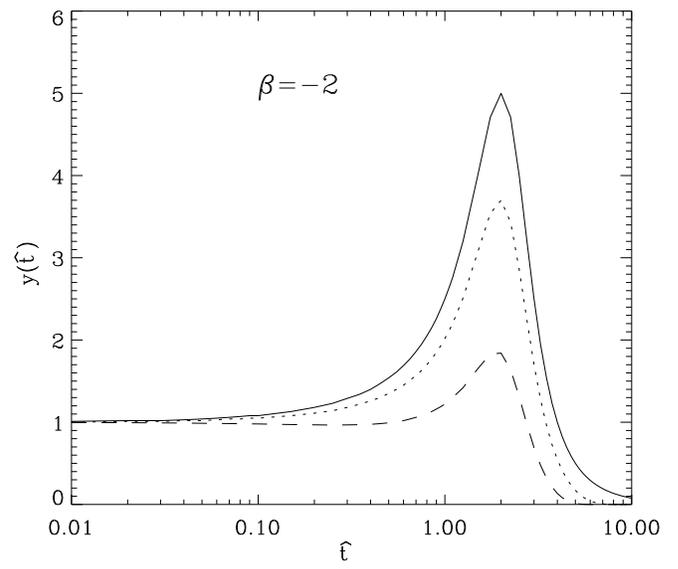
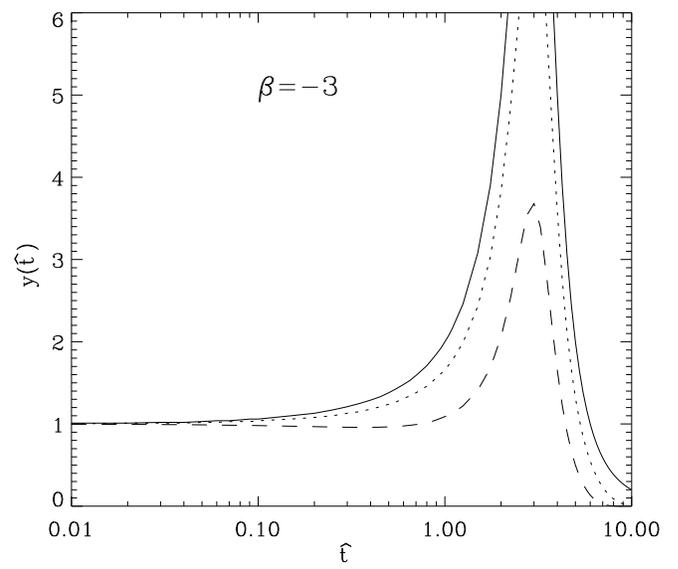
\end{document}